\pdfoutput=1

\documentclass[prl,twocolumn,showpacs,preprintnumbers,amssymb,superscriptaddress]{revtex4-1}
\usepackage{hyperref,graphicx,array,multirow,color,amsmath}
\hypersetup{pdftitle={},pdfcreator={},linkcolor=[rgb]{0.15,0.35,0.75},colorlinks=true,citecolor=[rgb]{0.675,0,0.2},urlcolor=[rgb]{0.15,0.35,0.65}}

\renewcommand{\bar}{\overline}
\newcommand{\fwbox}[2]{\text{\makebox[#1][c]{$\hspace{-150pt}\displaystyle#2\hspace{-150pt}$}}}
\newcommand{\fwboxL}[2]{\text{\makebox[#1][l]{$#2$}}}
\newcommand{\fwboxR}[2]{\text{\makebox[#1][r]{$#2$}}}
\renewcommand{\phi}{\varphi}
\newcommand{\eq}[1]{\vspace{-4.pt}\begin{equation}\hspace{-0pt}#1\hspace{-0pt}\vspace{-4.pt}\end{equation}}
\newcommand{\fig}[3]{\raisebox{#1}{\includegraphics[scale=#2]{#3}}}
\DeclareMathOperator*{\Res}{\mathrm{Res}}
\newcommand{\x}[2]{(#1,#2)}
\renewcommand{\u}{u_1}\renewcommand{\v}{u_2}\newcommand{\w}{u_3}
\newcommand{\mi}{\raisebox{0.75pt}{\scalebox{0.75}{$\hspace{-2pt}\,-\,\hspace{-0.5pt}$}}}
\renewcommand{\pl}{\raisebox{0.75pt}{\scalebox{0.75}{$\hspace{-2pt}\,+\,\hspace{-0.5pt}$}}}
\newcommand{\Zeta}[1]{\hspace{1pt}\zeta_{#1}}

\newcommand{\logk}[2]{\log^{\hspace{-0.25pt}#1}\hspace{-1pt}(#2)}
\newcommand{\weier}{Weierstra\ss~}
\newcommand{\usum}{\Sigma}
\newcommand{\uprod}{\Pi}
\newcommand{\toyint}{I_{\text{toy}}^{\text{ell}}}
\newcommand{\tenint}{I_{\text{db}}^{\text{ell}}}
\renewcommand{\alph}{{\color{hred}\alpha}}
\newcommand{\s}{{\color{hred}s}}
\newcommand{\norm}{\mathfrak{N}}
\newcommand{\fcn}[1]{G({\color{hblue}\hspace{0.5pt}\bar{\hspace{-0.75pt}w_{\color{hblue}i}\hspace{-3pt}}\hspace{2.5pt}},\hspace{-1pt}#1;\!\alph)}
\newcommand{\fcnb}[1]{G({\color{hblue}\hspace{0.5pt}\bar{\hspace{-0.75pt}w_{\color{hblue}i}\hspace{-3pt}}\hspace{2.5pt}};\!\alph)}
\newcommand{\wbar}[1]{\hspace{0.75pt}\bar{\hspace{-0.75pt}w_{#1}\hspace{-4pt}}\hspace{3pt}}
\newcommand{\wsumbar}{\hspace{1pt}\bar{\hspace{-1pt}w_1\pl w_2\hspace{-2pt}}\hspace{1pt}}
\definecolor{hblue}{rgb}{0,0,0.575}
\definecolor{hred}{rgb}{0.575,0.0,0.225}

\begin{document}
\title{\texorpdfstring{The Elliptic Double-Box Integral: Massless Amplitudes Beyond Polylogarithms\\[-18pt]~}{The Elliptic Double-Box Integral: Massless Amplitudes Beyond Polylogarithms}}
\author{Jacob~L.~Bourjaily}
\affiliation{Niels Bohr International Academy and Discovery Center, Niels Bohr Institute,\\University of Copenhagen, Blegdamsvej 17, DK-2100, Copenhagen \O, Denmark}
\author{Andrew~J.~McLeod}
\affiliation{Niels Bohr International Academy and Discovery Center, Niels Bohr Institute,\\University of Copenhagen, Blegdamsvej 17, DK-2100, Copenhagen \O, Denmark}
\author{Marcus~Spradlin}
\affiliation{Department of Physics, Brown University, 182 Hope Street, Providence, RI 02912, USA}
\affiliation{School of Natural Sciences, Institute for Advanced Study, 1 Einstein Drive, Princeton, NJ 08540, USA}
\author{Matt~von~Hippel}
\affiliation{Niels Bohr International Academy and Discovery Center, Niels Bohr Institute,\\University of Copenhagen, Blegdamsvej 17, DK-2100, Copenhagen \O, Denmark}
\author{Matthias~Wilhelm}
\affiliation{Niels Bohr International Academy and Discovery Center, Niels Bohr Institute,\\University of Copenhagen, Blegdamsvej 17, DK-2100, Copenhagen \O, Denmark}

\begin{abstract}
We derive an analytic representation of the ten-particle, two-loop double-box integral as an elliptic integral over weight-three polylogarithms. To obtain this form, we first derive a four-fold, rational (Feynman-)parametric representation for the integral, expressed {\it directly} in terms of dual-conformally invariant cross-ratios; from this, the desired form is easily obtained. The essential features of this integral are illustrated by means of a simplified toy model, and we attach the relevant expressions for both integrals in ancillary files. We propose a normalization for such integrals that renders all of their polylogarithmic degenerations pure, and we discuss the need for a new `symbology' of iterated elliptic/polylogarithmic integrals in order to bring them to a more canonical form.\\[-20pt]
\end{abstract}
\maketitle

\section{Introduction}\label{introduction_section}\vspace{-14pt}

In recent years, our ability to compute scattering amplitudes has advanced enormously. Loop {\it integrands} for scattering amplitudes are now known for a broad class of theories, loop orders, and multiplicities (see e.g.\ \mbox{\cite{ArkaniHamed:2010kv,ArkaniHamed:2010gh,Bourjaily:2015jna,Bourjaily:2015bpz,Bourjaily:2016evz,Bourjaily:2017wjl}}), and substantial inroads have been made towards the development of general loop integration technology \mbox{\cite{Smirnov:2006ry,Henn:2013pwa,Panzer:2014caa,conformalIntegration}}. Our understanding of the kinds of functions that result from these integrations has also experienced remarkable progress, especially in the case of (`Goncharov') hyperlogarithms \cite{goncharov:hyperlogs}, which capture most of perturbation theory at low orders and multiplicity~\cite{Goncharov:2010jf,Duhr:2012fh,Golden:2013xva,Dixon:2013eka,Duhr:2014woa,Caron-Huot:2016owq}. However, as exemplified by the two-dimensional sunrise integral with massive propagators (see e.g.\ \cite{Broadhurst:1993mw,Bloch:2013tra,Bloch:2014qca,Bloch:2016izu,Broadhurst:2016myo,Bogner:2017vim}), even the simplest quantum field theories are known to encounter elliptic and other non-polylogarithmic functions---for which the powerful tools of symbology, Hopf algebras, coproducts, etc.\ that have fueled such progress in the polylogarithmic case remain to be fully developed (but see e.g.\ \cite{Adams:2017ejb,Remiddi:2017har}).

In this work, we study what is perhaps the simplest non-polylogarithmic contribution to scattering amplitudes of massless particles in four dimensions: the elliptic double-box integral. This may be represented with either a Feynman diagram or its dual graph, depicted by\\[-10pt]~
\vspace{-4pt}\eq{\hspace{-60pt}\tenint\equiv\fwbox{82pt}{\fig{-22.25pt}{1.2}{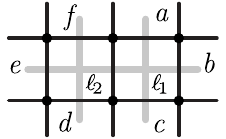}}\!=\fwbox{78pt}{\fig{-22.25pt}{1.2}{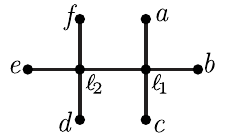}}\,.\label{double_box_figure}\hspace{-40pt}\vspace{-4pt}}
It may be viewed as a contribution to the ten-point amplitude in massless $\phi^4$ theory---but it also plays a significant role in (pure or supersymmetric) Yang-Mills and integrable fishnet theories \cite{Gurdogan:2015csr,Sieg:2016vap,Grabner:2017pgm}. In the context of planar maximally supersymmetric Yang-Mills, it is the sole diagram contributing to a particular helicity configuration \cite{CaronHuot:2012ab}, making it the entire amplitude in that case---and the same is true for the integrable fishnet theories. Considerations of maximal cuts and differential equations have led some authors to conjecture \cite{Paulos:2012nu,CaronHuot:2012ab,Nandan:2013ip,Chicherin:2017bxc} that (\ref{double_box_figure}) could be written schematically in the form
\vspace{-0pt}\eq{\tenint\sim\int\!\!\!\frac{d\alph}{\sqrt{Q(\alph)}}\big(\text{Li}_3(\cdots)\pl\ldots\big)\,,\label{schematic_target_form}\vspace{-0.75pt}}
where $Q(\alph)$ is an irreducible quartic in $\alph$, and thus encodes an elliptic curve. This form is attractive because it relates (\ref{double_box_figure}) to well-known functions while manifesting its ellipticity.

In this Letter, we realize such a representation explicitly by direct integration of Feynman parameters, without resorting to an ansatz or to solving differential equations. Specifically, we follow the strategy described in \mbox{ref.\ \cite{conformalIntegration}} to obtain a manifestly dual-conformally invariant, rational four-fold (Feynman-)parametric integral representation of $\tenint$, and carry out three of the integrations to obtain the desired form (\ref{schematic_target_form}). In what follows, we outline the steps involved, and describe how (\ref{schematic_target_form}) may be brought into a more canonical form with a normalization suggested by its degenerations. As we will see, this form points to the need for a `symbology' for mixed iterated elliptic/polylogarithmic integrals. For the sake of clarity and illustration, we first consider a simpler toy model of $\tenint$ restricted to a particular three-dimensional subspace of ten-particle kinematics that nevertheless preserves all of its essential structure. The full case of $\tenint$ will be described subsequently.

\vspace{-10pt}\section{Elliptic Toy Model}\label{elliptic_toy_model}\vspace{-14pt}

Our toy model depends symmetrically on only three cross-ratios. This is most directly described in terms of (the dual-momentum coordinates of) six massless particles, but it can also be obtained from $\tenint$ through a (maximal) sequence of constraints preserving ellipticity. 

\vspace{-10pt}\subsection{(Dual-Conformal) Loop Integration\texorpdfstring{\\}{ }via Feynman Parameterization}\label{loop_integration_toy_model}\vspace{-15pt}

In dual-momentum $x$-coordinates, the momentum of the $a^{\text{th}}$ external particle is defined as the difference \mbox{$p_a\!\equiv(x_{a+1}\mi x_{a})$} (with cyclic labeling understood).  In terms of these coordinates, we may define
\vspace{-2pt}\eq{\x{a}{b}\!=\!\x{b}{a}\!\equiv(x_{a}\mi x_b)^2\!=\!(p_a\pl\ldots\pl p_{b-1})^2.\vspace{-2pt}}
(`$\x{a}{b}$' is more frequently denoted `$x_{ab}^2$'.) Each loop momentum $\ell_i$ may be represented by a dual point $x_{\ell_i}$, and inverse propagators expressed as \mbox{$\x{\ell_i}{a}\!\equiv\!(x_{\ell_i}\mi x_{a})^2$}.

\newpage
Our toy model may be defined by taking the dual coordinates to describe the momenta of six massless particles by assigning $\{x_a,\ldots,x_f\}$ in (\ref{double_box_figure}) to $\{x_1,x_3,x_5,x_4,x_6,x_2\}$. That is, we impose the conditions
\vspace{-2pt}\eq{\fwbox{0pt}{(a,f)\!=\!(f,b)\!=\!(b,d)\!=\!(d,c)\!=\!(c,e)\!=\!(e,a)\!=\!0.}\vspace{-2pt}}
Note that, as these coordinates are assigned out of order, this choice does not correspond to a sensible massless limit of the diagram. While the resulting integral has no physical interpretation in terms of six-particle scattering, it does represent $\tenint$ evaluated on a well-defined three-dimensional subspace of ten-particle kinematics.

With this specialization, (\ref{double_box_figure}) can be written in dual coordinates as
\eq{\hspace{-40pt}\toyint\!\equiv\!\!\int\!\!\frac{\fwboxR{0pt}{d^4\ell_1d^4\ell_2\hspace{16pt}}\norm\,\x{1}{4}\x{2}{5}\x{3}{6}\fwboxL{0pt}{}}{\x{\ell_1}{1}\x{\ell_1}{3}\x{\ell_1}{5}\x{\ell_1}{\ell_2}\x{\ell_2}{4}\x{\ell_2}{6}\x{\ell_2}{2}}\,.\hspace{-30pt}\label{elliptic_hexagon_loop_space}}
We ignore overall numerical factors, but retain a kinematic-dependent normalization $\norm$ about which we will say more later. 
(Note that both $\toyint$ and $\tenint$ are finite, such that no regularization is required.)

We now transform (\ref{elliptic_hexagon_loop_space}) into a manifestly dual-conformally invariant \mbox{(Feynman-)}parametric integral. This is done by integrating one loop at a time, following the general strategy described in \mbox{ref.\ \cite{conformalIntegration}} (to which we refer the reader for more details). Using the embedding formalism (see e.g.\ \cite{Hodges:2010kq,SimmonsDuffin:2012uy}), we may associate Feynman parameters to the $\ell_1$ propagators according to
\vspace{-2pt}\eq{\hspace{-20pt}Y_1\!\equiv\!(1)\pl\beta_1(3)\pl\beta_2(5)\pl{\color{hblue}\gamma_1}(\ell_2)\!\equiv\!(R_1)\pl{\color{hblue}\gamma_1}(\ell_2)\,,\hspace{-20pt}\vspace{-2pt}}
where $(a)$ denotes the dual coordinate $x_a$. Letting $\mathcal{I}_{\text{toy}}^{\text{ell}}$ be the integrand of (\ref{elliptic_hexagon_loop_space}), the $\ell_1$ integration gives
\vspace{-4pt}\eq{\begin{split}\hspace{-100pt}\int\!\!d^4\ell_1\mathcal{I}_{\text{toy}}^{\text{ell}}&=\!\!\int\limits_0^\infty\!\!d^2\!\vec{\beta}\int\limits_0^\infty\!\!d{\color{hblue}\gamma_1}\frac{\norm\,\x{1}{4}\x{2}{5}\x{3}{6}\fwboxL{0pt}{}}{\x{Y_1}{Y_1}^2\x{\ell_2}{2}\x{\ell_2}{4}\x{\ell_2}{6}}\hspace{-30pt}\\
&\hspace{-0pt}=\!\!\int\limits_0^\infty\!\!d^2\!\vec{\beta}\frac{\norm\,\x{1}{4}\x{2}{5}\x{3}{6}\fwboxL{0pt}{\;}}{\x{R_1}{R_1}\x{\ell_2}{R_1}\x{\ell_2}{2}\x{\ell_2}{4}\x{\ell_2}{6}}\,,\hspace{-30pt}\\[-24pt]~\end{split}\label{elliptic_hexagon_gamma1_integration}}
where in the second line we have used the fact that the ${\color{hblue}\gamma_1}$ integral is a total derivative. For $\ell_2$, we introduce Feynman parameters according to
\eq{\hspace{-20pt}Y_2\!\equiv\!(R_1)\pl\alph(6)\pl\beta_3(2)\pl{\color{hblue}\gamma_2}(4)\!\equiv\!(R_2)\pl{\color{hblue}\gamma_2}(4)\,,\hspace{-20pt}}
and repeat the same steps as above (integrating out ${\color{hblue}\gamma_2}$), to obtain the four-fold representation
\vspace{-2pt}\eq{\toyint\!=\!\int\limits_0^{\infty}\!\!d\alph\!\!\int\limits_0^{\infty}\!\!d^3\!\vec{\beta}\frac{\norm\,\x{1}{4}\x{2}{5}\x{3}{6}\fwboxL{0pt}{}}{\x{R_1}{R_1}\x{R_2}{4}\x{R_2}{R_2}}\,.\label{elliptic_hexagon_pre_scaling}\vspace{-3pt}}

To render this {\it manifestly} dual-conformally invariant, we rescale the Feynman parameters according to
\vspace{-2pt}\eq{\hspace{-30pt}\alph\!\mapsto\!\alph\frac{\x{1}{3}}{\x{3}{6}},\,\beta_1\!\mapsto\!\beta_1\frac{\x{1}{5}}{\x{3}{5}},\,\beta_2\!\mapsto\!\beta_2\frac{\x{1}{3}}{\x{3}{5}},\,\beta_3\!\mapsto\!\beta_3\frac{\x{1}{5}}{\x{2}{5}},\hspace{-30pt}\nonumber\vspace{-2pt}}
after which (\ref{elliptic_hexagon_pre_scaling}) becomes simply
\vspace{-2pt}\eq{\hspace{-56pt}\toyint\!\equiv\!\!\int\limits_0^{\infty}\!\!\!d\alph\!\!\int\limits_0^{\infty}\!\!\!d^3\!\vec{\beta}\frac{\fwbox{0pt}{\norm}}{f_1f_2f_3},\left\{\hspace{-1pt}\begin{array}{@{}l@{}l@{}}f_1\!\equiv& \,\beta_1\pl\beta_2\pl\beta_1\beta_2\\f_2\!\equiv&\,1\pl\alph\hspace{1pt}\u\pl\v\beta_3\\f_3\!\equiv&\,f_1\pl\alph(\beta_1\pl\w\beta_3)\pl\beta_2\beta_3\end{array}\!\right\}\!\!.\hspace{-40pt}\label{toy_model_conformal_0}\vspace{1pt}}
This form depends {directly} on the familiar six-particle cross-ratios \mbox{$\u\!\equiv\!(13;\!46),\v\!\equiv\!(24;\!51),\w\!\equiv\!(35;\!62)$}, with
\vspace{-4pt}\eq{(ab;\!cd)\equiv\frac{\x{a}{b}\x{c}{d}}{\x{a}{c}\x{b}{d}}\,.\label{cross_ratio_notation_defined}\vspace{-3pt}}

To see that the integral (\ref{toy_model_conformal_0}) is elliptic (or at least non-polylogarithmic), it suffices to observe that
\eq{\Res_{f_i=0}\Big(\frac{d^3\vec{\beta}}{f_1f_2f_3}\Big)=\frac{1}{\sqrt{Q(\alph)}}\,,}
where $Q(\alph)$ is the irreducible quartic
\vspace{-1pt}\eq{\hspace{-50pt}Q(\alph)\!\equiv\!(1\pl\alph(\u\pl\v\pl\w\pl\alph\hspace{1pt}\u\w))^2\mi4\alph(1\pl\alph\hspace{1pt}\u)^2\w.\hspace{-30pt}\label{qalph_of_toy}\vspace{-1pt}}

The $\beta_i$ integrals of (\ref{toy_model_conformal_0}) can be done analytically using standard methods (e.g.\ using \cite{Panzer:2014caa}). Doing so results in
\vspace{-2pt}\eq{\toyint=\int\limits_0^{\infty}\!\!d\alph\frac{\norm}{\sqrt{Q(\alph)}}H_{\text{toy}}(\alph)\,,\label{direct_form_after_partial_integration}\vspace{-2pt}}
where $H_{\text{toy}}(\alph)$ is a sum of pure weight-three hyperlogarithms that depend on the final integration parameter. Explicitly, this function may be written in terms of \mbox{$H_{\text{toy}}(\alph)\equiv F_1(\alph)\mi F_2(\alph)$}, where
\begin{widetext}\vspace{-16pt}\begin{align}\hspace{-7pt}F_{\color{hblue}i}(\alph)\!\equiv&\phantom{\pl}\fcn{0,0}\pl\fcn{\overline{0},\bar{0}}\mi\fcn{0,\bar{0}}\mi\fcn{\bar{0},0}\mi\fcn{\mi\wbar{1}\wbar{2},0}\mi\fcn{\frac{\wbar{1}\wbar{2}}{\wsumbar},\bar{0}}\nonumber\\
&\pl\fcn{\frac{\wbar{1}\wbar{2}}{\wsumbar}}\log(w_1w_2\wbar{1}\wbar{2})\mi\fcn{\mi\wbar{1}\wbar{2}}\log\!\Big(\hspace{-1pt}\frac{\mi1}{\wbar{1}\wbar{2}}\Big)\pl\big(\hspace{-1pt}\fcn{0}\mi\fcn{\bar{0}}\hspace{-1pt}\big)\!\log(\!\mi w_1w_2)\label{explicit_halpha_formula_toy}\\
&\pl\fcnb{}\hspace{-3pt}\left\{\!\!{\textstyle\frac{1}{2}}\!\log^{\hspace{-0.25pt}2}\hspace{-2pt}\Big(\hspace{-1pt}\frac{1}{\wsumbar}\Big)\pl\log(w_1w_2)\!\log\!\Big(\hspace{-1pt}\frac{\mi1}{\wbar{1}\wbar{2}}\Big)\mi\!\log\!\Big(\hspace{-1pt}\frac{1}{\wsumbar}\Big)\!\log\!\Big(\hspace{-1pt}\frac{1}{\wbar{1}\wbar{2}}\Big)\pl\text{Li}_2\Big(\mi\frac{\wsumbar}{\wbar{1}\wbar{2}}\Big)\hspace{-2pt}\right\}\!,\nonumber\\[-20pt]~\nonumber
\end{align}\end{widetext}
where the short-hand $\bar{x}\!\equiv\!\mi1/(1\pl x)$ (so that $\bar{0}\!=\!\mi1$), \mbox{$\fcn{\ldots}$} is an ordinary Goncharov polylogarithm \cite{goncharov:hyperlogs}, and \mbox{$w_{{\color{hblue}1},{\color{hblue}2}}\!\equiv\!\Big[\alph\big((\alph\hspace{0.5pt}\w\mi1)\u\mi\v\pl\w\big)\mi1\!{\color{hblue}\,\pm\,}\!\sqrt{Q(\alph)}\Big]/(2\alph\hspace{0.5pt}\v)$}. This form of $H_{\text{toy}}(\alph)$ is not manifestly real, but we have been careful in our expression above to ensure that the imaginary parts cancel for sufficiently canonical branch choices---e.g.\ the defaults chosen by {\tt GiNaC} \cite{Vollinga:2004sn}. 

In the ancillary files to this Letter, we have included an expression for $H_{\text{toy}}(\alph)$ in terms of classical polylogarithms (which is always possible at this weight \cite{Goncharov:motivicGalois}) that are manifestly real along the entire contour of integration $\alph\!\in\![0,\infty]$. As such, we have realized an expression of this toy model in the conjectured form (\ref{schematic_target_form}). However, this representation is still far from unique---even after choosing a basis of hyperlogarithms. This is partially due to a freedom to re-parametrize the quartic in the integration measure. This redundancy can by resolved by bringing the elliptic curve encoded by the quartic into a standard (e.g.\ \weier\!) form, which we now describe.

\vspace{-18pt}\subsection{Toward Canonicalization (via \texorpdfstring{\weier\!}{Weierstrass}) }\label{canonical}\vspace{-15pt}

One of the advantages of working with hyperlogarithms is that all polylogarithmic identities are enforced within a given choice of fibration basis~\cite{Panzer:2014caa}. In seeking a `canonical' form for integrals of the form (\ref{schematic_target_form}), we thus hope to realize similar advantages. There are at least three desiderata one might seek for a preferred representation of such integrals: the representation should (i) provide a prescription that determines the normalization $\norm$ and fixes the parametrization of the quartic, (ii) automatically enforce all functional identities, and (iii) make manifest any symmetries that are respected by the full integral. The integral representation of $\toyint$ derived above does not automatically meet any of these criteria: the elliptic curve encoded by $Q(\alph)$ may be parametrized in many ways, there may exist nontrivial identities between integrals of this form, and $\toyint$ has symmetries that are not manifest in (\ref{direct_form_after_partial_integration}). (In particular, the loop-momentum-space definition of $\toyint$ in (\ref{elliptic_hexagon_loop_space}) is fully permutation invariant in $\{\u,\v,\w\}$; but this is obscured in both the Feynman-parametrized integral and subsequent integrated expression: neither $Q(\alph)$ in (\ref{qalph_of_toy}) nor the hyperlogarithms that result from $\beta_i$ integrations are permutation invariant.)

All the symmetries of $\toyint$ can be made manifest  at least in the integration measure by transforming it into \weier form. This is accomplished by a standard map $\alph\!\mapsto\!f(\s, \{ u_i \})$ such that 
\vspace{-1pt}\eq{\hspace{-46pt}{Q(\alph)}\!\mapsto\!{Q(\s)}\!\equiv\!{4\s^3\mi g_2\s\mi g_3}\!\equiv\!{4(\s\mi e_1)(\s\mi e_2)(\s\mi e_3)}\,,\hspace{-30pt}\label{weier_quartic}\vspace{-1pt}}
after which (\ref{direct_form_after_partial_integration}) becomes
\vspace{-4pt}\eq{\toyint\equiv\int\limits^\infty_{\fwbox{6pt}{\usum^2\!/3}}\!\!d\s\frac{2\,\norm\,}{\sqrt{4\s^3\mi g_2\s\mi g_3}}H_{\text{toy}}(\s)\,,\label{weier_form_of_toy_v1}\vspace{-5pt}}
where \mbox{$\usum\!\equiv\!(\u\pl\v\pl\w)$}, \mbox{$\uprod\!\equiv\!\u\v\w$}, and 
\vspace{-1pt}\eq{\hspace{-46pt}g_2\equiv\frac{4}{3}\Big(\usum^4\mi24 \uprod\usum\Big),\; g_3\equiv\frac{32}{3}\Big(\uprod(\usum^3\mi6\uprod)\mi\frac{1}{36}\usum^6\Big).\hspace{-30pt}\label{weier_gs_for_toy}\vspace{-1pt}}
The (elliptic) integration measure is now {\it manifestly} symmetric in the cross-ratios. 

The modular discriminant $\Delta$ is given by
\vspace{-2pt}\eq{\Delta\!\equiv\!g_2^3\mi27g_3^2\!=\!(16\uprod)^3(\usum^3\mi27\uprod)\,.\label{toy_discriminant}\vspace{-2pt}}
So long as $\Delta\!>\!0$, the roots of the cubic $e_i$ in (\ref{weier_quartic}) will be real. It is standard to order them $e_1\!>\!e_2\!>\!e_3$ so that the modulus $k\!\equiv\!\sqrt{(e_2\mi e_3)/(e_1\mi e_3)}$ is also manifestly real. $\Delta\!>\!0$ is the kinematic domain in which the integral (\ref{weier_form_of_toy_v1}) is defined. It is not hard to see that this corresponds to the entire Euclidean domain ($u_i\!>\!0$) {\it except} along the line \mbox{$\u\!=\!\v\!=\!\w$}. 

The analytic form of $H_{\text{toy}}(\s)$ can be obtained from $H_{\text{toy}}(\alph)$ by direct substitution (being careful to account for the implicit dependence of $\alph$ in $w_i$). Importantly, even putting $\toyint$ into \weier form, the function $H_{\text{toy}}(\s)$ is {\it still not} automatically permutation invariant! This points to the existence of identities between mixed elliptic/polylogarithmic integrals that are still not accounted for. We expect that eliminating such redundancies will require the development some analogue of `symbology' for mixed integrals of these types, perhaps along the lines of \mbox{refs.\ \cite{Adams:2017ejb,Remiddi:2017har,Ablinger:2017bjx}}.

The \weier map is thus not sufficient to achieve desiderata (ii) or (iii). (It is true that $H_{\text{toy}}(s)$ can be put in a form that respects (iii) by appropriately summing over its permutations, but this would merely obfuscate the underlying issue, whose resolution requires a deeper understanding of these types of integrals.) However, let us now turn to the remaining issue (i) raised above: how these integrals should be normalized.

\vspace{-18pt}\subsection{Normalization: A Proposal for Elliptic `Purity'}\label{normalization}\vspace{-15pt}

Let us now discuss how the integral (\ref{weier_form_of_toy_v1}) should be normalized by considerations of `purity'. For an integral with only logarithmic singularities (locally expressible everywhere in the form $\prod_id\log(\alpha_i)$, \cite{Arkani-Hamed:2014via}), purity simply means that all its maximal co-dimension residues have unit magnitude. All hyperlogarithms are pure by definition. When an integral has {\it no} residues with maximal co-dimension, such as the integrals studied in this Letter, it is {\it a priori} unclear what `purity' should mean. This is the reason we have allowed for an unknown normalization $\norm$ in our integral (\ref{elliptic_hexagon_loop_space}). It may turn out that the right notion of a `pure' mixed elliptic/polylogarithmic function will require a better understanding of their co-product structure, but a candidate for this normalization follows naturally from degenerate limits where the integral becomes polylogarithmic. 

To examine a degenerate limit in which the integral (\ref{toy_model_conformal_0}) has maximal co-dimension residues we consider the \weier form (\ref{weier_form_of_toy_v1}), where this happens if and only if two of the roots $e_i$ in (\ref{weier_quartic}) collide. When these roots are real and canonically ordered, only $\{e_1,e_2\}$ or $\{e_2,e_3\}$ may become degenerate---$(e_1\mi e_3)$ is always positive. More geometrically, the degeneration of the elliptic curve would be signaled by the modulus $k$ approaching $1$ or $0$. In this case, the $\alph$ integration {\it does indeed} have poles: for example if $e_2\!=\!e_1$,
\eq{\Res_{\s=e_1}\Big(\frac{d\s\,\norm}{\sqrt{(\s\mi e_1)^2(\s\mi e_3)}}\Big)=\frac{\norm}{\sqrt{e_1\mi e_3}}\,.}
This clearly shows that if we normalize the original integral by taking $\norm\!\equiv\!\sqrt{e_1\mi e_3}$, then at least all polylogarithmic degenerations will automatically be pure in the conventional sense. As this normalization renders {\it all} degenerations manifestly pure polylogarithmic iterated integrals, we propose that this be considered a canonical choice. Thus, the toy model integral should be written with the normalization
\vspace{-1.5pt}\eq{\hspace{-50pt}\toyint\equiv\int\limits^\infty_{\fwbox{6pt}{\usum^2\!/3}}\!\!d\s\frac{\sqrt{e_1\mi e_3}}{\sqrt{(\s\mi e_1)(\s\mi e_2)(\s\mi e_3)}}H_{\text{toy}}(\s)\,.\hspace{-50pt}\label{weier_form_of_toy_final}\vspace{-3pt}}

Another motivation for discussing the {\it residues} of $\toyint$ in these degenerate limits (opposed to the functional limits themselves) is the fact that, for this toy model, all degenerations are infrared-divergent. (This will not be the case for the double-box integral $\tenint$, which admits finite non-elliptic limits.) The degenerations of $\toyint$ correspond to $\Delta\!\to\!0$ in (\ref{toy_discriminant}); this occurs, for example, when any $u_i\!\to\!0$. Such degenerations can generally be represented analytically as series expansions in $\log(u_i)$ \mbox{\cite{conformalIntegration}}. For example,
\vspace{-0pt}\eq{\hspace{-70pt}\lim_{{\color{hblue}\w}\to0}\!\!\Big(\!\toyint\!\Big)\!\!\propto\!\!\frac{\sqrt{e_1\mi e_3}}{\u\pl\v}\Big[\!\logk{2}{{\color{hblue}\w}}\Big(\!\!\log^{\hspace{-0.25pt}2}\hspace{-1pt}\Big(\frac{\u}{\v}\Big)\pl6\Zeta{2}\Big)\pl\!\ldots\!\Big],\hspace{-50pt}\vspace{-0pt}}
where the additional terms are those less divergent as ${\color{hblue}\w}\!\to\!0$. Because $\uprod\!\to\!0$ in this limit, it is easy to see that $\!\displaystyle\lim_{{\color{hblue}\w}\to0}(\sqrt{e_1\mi e_3})\!=\!(\u\pl\v)$, rendering the limit pure.

\vspace{-12pt}\section{Elliptic Double-Box Integral}\label{elliptic_double_box}\vspace{-14pt}

Let us now turn our attention to the actual elliptic double-box integral $\tenint$ shown in (\ref{double_box_figure}). In dual-momentum coordinates, this integral may be written as
\eq{\hspace{-70pt}\tenint\!\equiv\!\int\!\!\!\frac{\fwboxR{0pt}{d^4\ell_1d^4\ell_2\hspace{10pt}}\norm\,\x{a}{c}\x{b}{e}\x{d}{f}}{\x{\ell_1}{a}\x{\ell_1}{b}\x{\ell_1}{c}\x{\ell_1}{\ell_2}\x{\ell_2}{d}\x{\ell_2}{e}\x{\ell_2}{f}}\,,\hspace{-50pt}\label{loop_space_double_box}}
where the pairs of dual points $\{x_a,x_f\}$ and $\{x_c,x_d\}$ are understood to be null-separated: \mbox{$\x{a}{f}\!=\!\x{c}{d}\!=\!0$}. 

Following the same sequence of Feynman parameterizations and loop-integrations as before---explicitly, using\\[-8pt]
\vspace{-0pt}\eq{\begin{split}~\\[-20pt]Y_1&\equiv(b)\pl\alph(c)\pl\beta_1(a)\pl{\color{hblue}\gamma_1}(\ell_2)\equiv (R_1)\pl{\color{hblue}\gamma_1}(\ell_2)\,,\\
Y_2&\equiv(R_1)\pl\beta_2(f)\pl\beta_3(d)\pl{\color{hblue}\gamma_2}(e)\equiv(R_2)\pl{\color{hblue}\gamma_2}(e)\,,\hspace{-5pt}\\[-20pt]~\end{split}\vspace{-0pt}}
to parametrize the propagators and recognizing the ${\color{hblue}\gamma_i}$ integrations as total derivatives---we arrive at an expression quite similar to (\ref{elliptic_hexagon_pre_scaling}):
\eq{\tenint\!=\!\int\limits_0^{\infty}\!\!d\alph\!\!\int\limits_0^{\infty}\!\!d^3\!\vec{\beta}\frac{\norm\,\x{a}{c}\x{b}{e}\x{d}{f}\fwboxL{0pt}{}}{\x{R_1}{R_1}\x{R_2}{e}\x{R_2}{R_2}}\,.\label{double_box_pre_scaling}}
Upon rescaling the Feynman parameters according to
\eq{\hspace{-50pt}\alph\!\mapsto\!\alph\frac{\x{a}{b}}{\x{a}{c}},\,\beta_1\!\!\mapsto\!\beta_1\frac{\x{b}{c}}{\x{a}{c}},\,\beta_2\!\!\mapsto\!\beta_2\frac{\x{b}{d}}{\x{d}{f}},\,\beta_3\!\!\mapsto\!\beta_3\frac{\x{b}{f}}{\x{d}{f}},\nonumber\hspace{-45pt}}
we obtain the following dual-conformally invariant expression:
\vspace{-2pt}\eq{\hspace{-40pt}\tenint\!=\!\int\limits_0^{\infty}\!\!d\alph\!\!\int\limits_0^{\infty}\!\!d^3\!\vec{\beta}\frac{\norm}{f_1\,f_2\,f_3}\,,\hspace{-40pt}\vspace{-6pt}\label{double_box_four_fold_Feynman}\vspace{-2pt}}
where
\eq{\hspace{-100pt}\begin{array}{l}f_1\equiv\alph(1\pl\beta_1)\pl\beta_1,\;f_2\equiv1\pl u_1\alph\pl v_1\beta_1\pl u_2\beta_2\pl v_2\beta_3,\\
f_3\equiv (1\pl u_3\alph)\beta_2\pl(1\pl u_4\beta_1)\beta_3\pl\beta_2\beta_3\pl u_3u_4u_5 f_1,\end{array}\hspace{-80pt}}
which depend on the seven dual-conformal cross-ratios
\eq{\hspace{-102pt}\begin{array}{l@{$\;$}l@{$\;$}l}u_1\!\equiv\!(ab;\!ce),&u_2\!\equiv\!(bd;\!ef),&u_3\!\equiv\!(ab;\!cf),\\
\fwbox{10.275pt}{v_1}\!\equiv\!(ea;\!bc),&\fwbox{10.275pt}{v_2}\!\equiv\!(fb;\!de),&u_4\!\equiv\!(bc;\!da),\end{array}u_5\!\equiv\!(ac;\!df).\hspace{-80pt}\label{ten_point_cross_ratios}}

As before, the $\beta_i$ integrals can be done analytically to give weight-three hyperlogarithms that depend on the final integration variable. This results in a representation of the form
\vspace{-2pt}\eq{\tenint=\int\limits_0^{\infty}\!\!d\alph\frac{\norm}{\sqrt{Q(\alph)}}H(\alph)\,,\label{elliptic_after_partial_integration}\vspace{-2pt}}
where
\eq{\hspace{-70pt}\begin{array}{@{}l@{}}Q(\alph)\equiv\big((\alph(u_4\mi1)\mi1)u_2\pl h_1\pl h_2\big)^2-4 h_1 h_2,\,\text{with}\\
h_1\equiv (1\pl\alph)(1\pl\alph u_3)v_2,\,\,\,\,\hspace{0.65pt}h_2\equiv 1\pl\alph(1\pl(1\pl\alph)u_1\mi v_1)\end{array}\hspace{-50pt}\label{qalph_for_tenint}}
is an irreducible quartic. While this is schematically of the desired form (\ref{schematic_target_form}), we again prefer to map it to \weier form to make manifest the symmetries of the full integral in the quartic (at least), and to normalize it according to our above prescription. 

The elliptic double-box integral is symmetric under two reflections. Written in dual-momentum coordinates, these correspond to \mbox{$r_1\!:\!\{a,\!b,\!c,\!d,\!e,\!f\}\!\mapsto\!\{c,\!b,\!a,\!f,\!e,\!d\}$} and \mbox{$r_2\!:\!\{a,\!b,\!c,\!d,\!e,\!f\}\!\mapsto\!\{f,\!e,\!d,\!c,\!b,\!a\}$}. The first of these merely permutes the cross-ratios defined in (\ref{ten_point_cross_ratios}) via \mbox{$r_1\!:\!\{u_1,\!v_1,\!u_2,\!v_2,\!u_3,\!u_4,\!u_5\}\!\mapsto\!\{v_1,\!u_1,\!v_2,\!u_2,\!u_4,\!u_3,\!u_5\}$}, while the second acts somewhat less trivially:
\vspace{-2.5pt}\eq{r_2\!:\!\{u_1,\!v_1,\!u_2,\!v_2,\!u_3,\!u_4,\!u_5\}\!\mapsto\!\{u_2,\!v_2,\!u_1,\!v_1,\!\frac{u_4u_2}{v_1},\!\frac{u_3v_2}{u_1},\!u_5\}.\nonumber\vspace{-2.5pt}}
The quartic $Q(\alph)$ in (\ref{qalph_for_tenint}) possesses neither of these symmetries; but as before, they become manifest once it is brought into \weier form via (\ref{weier_quartic}). This gives rise to the integral representation
\vspace{-2pt}\eq{\tenint\equiv\int\limits^\infty_{\fwbox{6pt}{\s_0}}\!\!d\s\frac{\sqrt{e_1\mi e_3}}{\sqrt{(\s\mi e_1)(\s\mi e_2)(\s\mi e_3)}}H(\s)\,, \label{weier_form_of_db_final}\vspace{-2pt}}
where $\s_0$ is the image of $\alph\!=\!\infty$ under the transformation to \weier form. Notably, $\s_0$ does \emph{not} respect the same permutation symmetries as $\tenint$. Thus, it is not possible to bring $H(\s)$ into a form that respects the symmetries of the full integral (under a single integration sign). The fact that our chosen normalization renders all polylogarithmic degenerations pure in the conventional sense is much less trivial in this case than in the toy model. This integral (\ref{weier_form_of_db_final}) admits many polylogarithmic limits (as well as one to $\toyint$). For example, when $x_f\!\to\!x_a$ or $x_c\!\to\!x_d$, the integral becomes polylogarithmic (and still infrared-finite). The appropriate normalizations of these limits are quite different, but the normalization in (\ref{weier_form_of_db_final}) ensures the purity of them all.

In the ancillary files for the Letter, we give an expression for $H(\alph)$ in terms of classical polylogarithms---valid throughout the `positive' part of the Euclidean domain. 

\newpage
\vspace{-14pt}\section{Conclusions and Outlook}\label{conclusions_and_outlook}\vspace{-14pt}

We have shown that straightforward Feynman parameterization and integration can be carried out for the elliptic double-box integral, resulting in a manifestly dual-conformally invariant representation as an integral over a standardized elliptic measure times a weight-three hyperlogarithm. Nevertheless, even after both parts of the integrand have been separately put into canonical forms, there exist non-trivial functional identities. Thus, our work emphasizes the need for a better understanding of `symbology' relevant to such cases. We expect that converting our results into iterated integrals over modular forms (as suggested in \cite{Adams:2017ejb}) will help, but we leave this to future work. 

Finally, we should point out that there is a curious (if not fully established) correspondence between Feynman integrals with external masses and massless propagators and those with massless external particles and massive propagators. Thus, we expect that our work may have some relevance to the more phenomenologically-motivated cases studied in \mbox{e.g.\ ref.\ \cite{Sogaard:2014jla}}.

\vspace{15pt}\acknowledgments
We gratefully acknowledge useful conversations with Lauren Altman, David Broadhurst, Nima Arkani-Hamed, Jaroslav Trnka, Anastasia Volovich, and Ellis Yuan, and the hospitality of the Institute for Advanced Study in Princeton. We thank Jaroslav Trnka for suggesting the toy model. This work was supported in part by: the US Department of Energy under contract DE-SC0010010 Task A (MS), the Simons Fellowship Program in Theoretical Physics (MS), the Danish Independent Research Fund under grant number DFF-4002-00037 (MW), the Danish National Research Foundation under grant number DNRF91 and the Villum Fonden  (JLB,AJM,MvH,MW).

\vspace{-15pt}
\providecommand{\href}[2]{#2}\begingroup
\endgroup
\end{document}